\documentclass[12pt]{article}
\usepackage{lineno}
\usepackage[margin=1in]{geometry}
\usepackage{amssymb,amsmath,latexsym}                        
\usepackage{graphicx}
\usepackage{cite}
\usepackage{xcolor}
\usepackage{float}

\usepackage{amsthm}
\usepackage{multicol}
\usepackage{authblk}

\usepackage[colorlinks]{hyperref}

\usepackage{adjustbox}
\usepackage{bbm}
\usepackage{mathtools}

\usepackage{graphicx}
\usepackage{subcaption}
\date{}

\begin{document}
\title{Improving the spatial resolution of NICA/MPD ECAL with new reconstruction methods}
\author[1,2]{Fuyue Wang}
\author[1,2]{Dong Han}
\author[1,2]{Yi Wang\footnote{Corresponding author. Email: yiwang@mail.tsinghua.edu.cn.}}
\author[1,2]{Chendi Shen}
\author[1,2]{Yulei Li}
\author[3]{Igor Tyapkin}
\author[3]{Dabrowska Boyana Rumenova}
\author[1,2]{Yuanjing Li}
\affil[1]{\normalsize\it Department of Engineering Physics, Tsinghua University, Beijing 100084, China}
\affil[2]{\normalsize\it Key Laboratory of Particle and Radiation Imaging(Tsinghua University), Ministry of Education, Beijing 100084, China}
\affil[3]{Joint Institute for Nuclear Research, JINR, Joint-Curie 6, 141980 Dubna, Moscow region, Russia}

\maketitle
\begin{abstract}

A Shashlyk-type electromagnetic calorimeter (ECal) will be used in the Multi-purpose Detector at Nuclotron-based Ion Collider facility to study the properties of nuclear matter. In this experiment, the ECal detector is responsible for measuring the energy and position of the incident particles, and identifying them with the information obtained from itself and other detectors. This paper analyzes the position resolution of the Tsinghua ECal modules using the data from a beam test in DESY. Several reconstruction methods are studied in detail. With a deep learning based algorithm, the position resolution of the prototypes achieves less than 3.8 mm for 1.6 GeV electron beam, which is improved by about 30\% compared to that of traditional charge center of gravity method. 
\end{abstract}

%------------------------------------------------------
\section{Introduction}
\label{sec:intro}
The electromagnetic calorimeter (ECal) is an important component of the Multi-purpose Detector (MPD) at Nuclotron-based Ion Collider fAcility (NICA) in Joint Institute for Nuclear Research (JINR), Dubna. The NICA collider is designed to provide collisions of heavy-ions over a wide range of atomic masses (from proton to Au) and operate at a luminosity of up to $L=10^{27} \rm cm^{-2}s^{-1}$ for $Au+Au$ collisions. The center-of-mass energy is $\sqrt{s_{NN}}=4\sim11 \rm GeV$\cite{abraamyan2011mpd}. The MPD is located at one of the two interaction points on NICA and is a $4\pi$ spectrometer capable of detecting charged hadrons, electrons and photons. This experiment is dedicated to study hot and dense strongly interacting QCD matter and search for a possible manifestation of the mixed phase formation and critical endpoint in heavy ion collisions\cite{golovatyuk2016multi}.

The main goals of the electromagnetic calorimeter in the MPD detectors are particle identification, photon flux measurement, and reconstruction of decays with photons involved\cite{abraamyan2011mpd}. These impose certain requirements for the detector's resolution on energy, time and position, as well as the granularity, occupancy and so on. A good positioning ability of ECal guarantees an effective separation of the overlapping showers and a precise reconstruction of the position, which is very important for both the charged and neutral particles. Because for charged ones, their signals in ECal are matched with the trajectories reconstructed from the tracking detector in order to make accurate particle identification. As for the neutral particles, especially photons that are important to reconstruct $\pi^0$, the position measurement is only available from the calorimeter.

The position of the incident particle is usually obtained using the method of Charge Center of Gravity (CCOG) with some additional corrections. This method calculates the average position of the related towers weighted by charge, and is a traditional algorithm in many position sensitive detectors. However, detailed studies on this method and the corrections for the ECal detector are rare, let alone new reconstruction methods. In this article, the CCOG method accompanied with different corrections are studied in order to improve the resolution. Meanwhile, a new method based on the neural networks is proposed, and the results are proved to be even better.

The paper is organized as follows: Sec.\ref{sec:experiment} describes the structure of the Ecal prototypes and the experiment setup in DESY. Sec.\ref{sec:method} describes the methods used to reconstruct the position. The charge center of gravity with multiple corrections is studied in detail in Sec.\ref{sec:inter}, while a new algorithm based on the neural network and deep learning is proposed in Sec.\ref{sec:NN}. Sec.\ref{sec:results} compares the results given by all the methods mentioned in this paper. Finally, Sec.\ref{sec:conclu} concludes the paper.

\begin{figure*}[h!]
	\centering
	\includegraphics[width=0.7\textwidth]{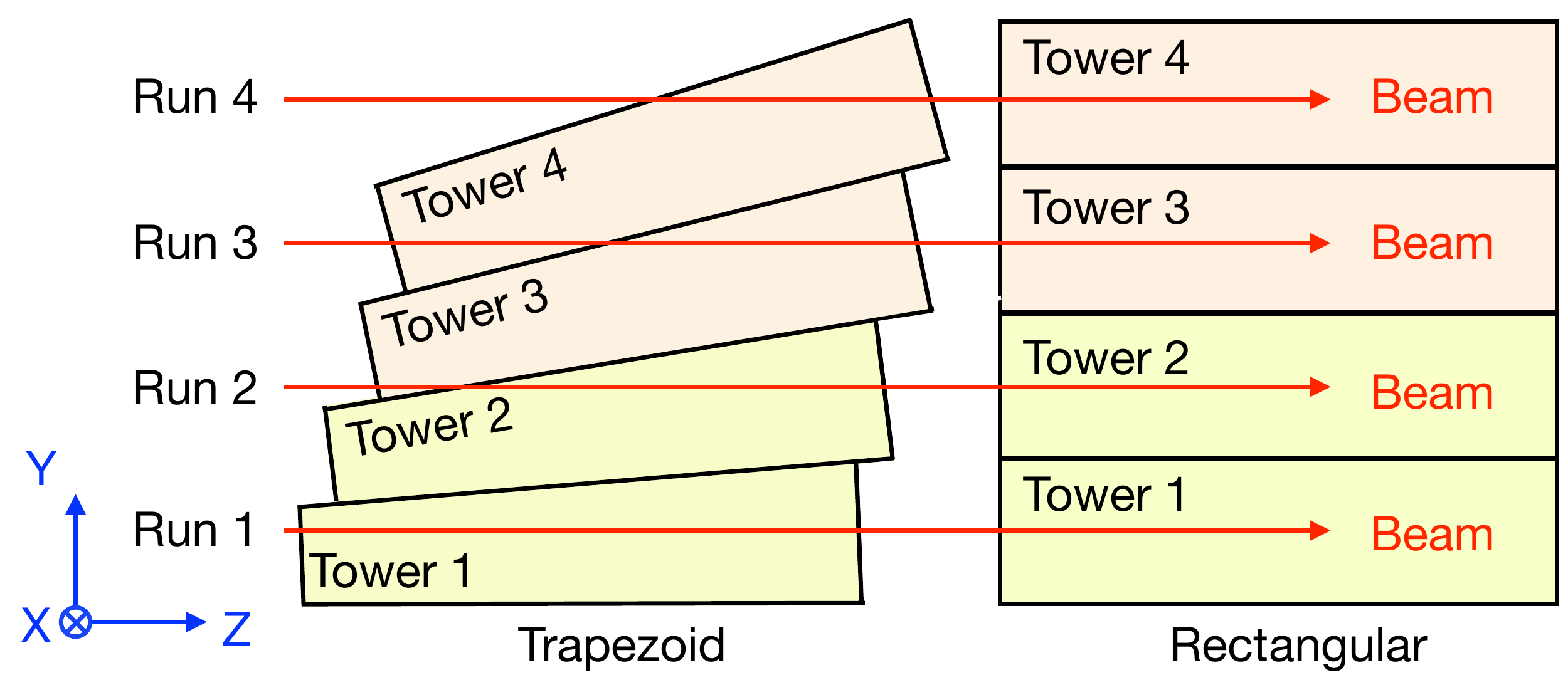}
	\caption{Two geometries of the electromagnetic calorimeter. The left configuration shows the trapezoid towers, while the right are rectangular ones. The red lines in the figure represent the positional relationships between the beam and the detector in 4 different experiment runs.}
	\label{fig:geo}
\end{figure*} 

\section{Experiment setup}
\label{sec:experiment}
The ECal prototype built in Tsinghua University is a Pb-scintillator sampling calorimeter of the shashlyk-type. Each prototype consists of 16 towers arranged in 2 rows (Y direction) and 8 columns (X direction). These towers are disposed together in the way of the left configuration in Fig.\ref{fig:geo}. Towers are basic building elements of the prototype, and each of them contains 220 alternating tiles of Pb (0.3 mm) and plastic scintillators (1.5 mm). Layers in each tower are optically combined by 16 longitudinally penetrating wavelength shifting fibers (WLS). These fibers are meant to collect the scintillation light, and the light is read out by the silicon photomultiplier (SiPM). The total thickness of the tower is about 40 $\rm cm$, which corresponds to 12 radiation lengths. Every tower is cut from 4 sides with an angle of 0.5 degree with respect to the axial direction. After such a cut, towers have a trapezoid shape in the $r\phi$ plane and this will significantly reduce the dead zones effect compared to the rectangular ones. More specific parameters of the ECal prototype can be found in \cite{nicatdr}. The comparison of the two geometries is presented in Fig.\ref{fig:geo}. In each configuration, two prototypes are placed one on the top of the other (they are in different colors), and each of them has two rows of towers. The left configuration shows the trapezoid towers, while the right are rectangular ones. The pitch size of the rectangular tower is $4\times4$ $\rm cm^2$, while for the trapezoid one, 4 cm is the size of the larger base. The red lines in the figure represent the relative positional relationships between the beam and the detector in 4 different experiment runs.

\begin{figure*}[h!]
	\centering
	\includegraphics[width=0.8\textwidth]{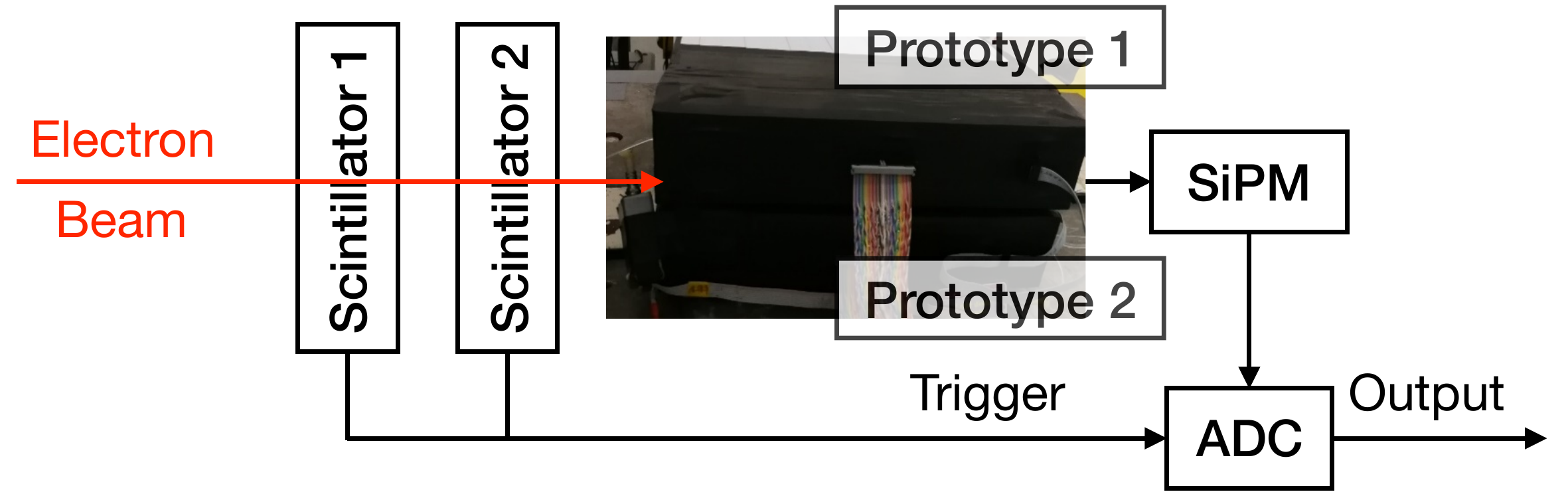}
	\caption{Experiment setup.}
	\label{fig:setup}
\end{figure*} 

In August 2018, two ECal prototypes were tested at Deutsches Elektronen-Synchrotron(DESY) using an electron beam. This is a converted bremsstrahlung beam with energy ranging from 1 to 6 GeV. Considering the size of the trigger counter and the angular distribution, the uncertainty of the beam position is 3 $\rm mm$. Fig.\ref{fig:setup} presents the setup of the beam test. 32 towers in 2 prototypes were readout separately by 32 SiPMs. These SiPM channels were then connected to an ADC to read out the signal waveforms. The sampling rate of the ADC is 0.0625Gs/s, and about 60 points are recorded for an entire waveform. An LED with short blue light was used to calibrate the energy in different channels, and the coincidence of the 2 scintillators provided triggers for the beam test.

In order to study the position resolution, 2 prototypes shown in Fig.\ref{fig:geo} were placed on a movable platform whose position can be adjusted remotely and measured by an accurate position finder. The resolution of this finder is 0.1 $\rm mm$, and the "truth" position of ECal is given by it. The detector was moved along the X and Y directions respectively, each time with a small step (on the order of millimeters). Fig.\ref{fig:geo} displays 4 examples of the experiment runs during the scanning in Y direction, and over 400,000 events were collected in each run. The center of the corresponding tower was kept aligned with the beam position in the X (Y) direction while the detector moved along Y (X). The scanning of the position was also performed at different energies ranging from 1.0 to 2.2 GeV, so as to study the relationship between the position resolution and the beam energy. Scannings in towers of the middle 2 rows and several middle columns are analyzed in this paper, since the information of the scannings in edge towers is limited by the geometry.

\section{Method of position reconstruction}
\label{sec:method}
\subsection{Charge center of gravity with corrections}
\label{sec:inter}
\begin{figure*}[h!]
	\centering
	\includegraphics[width=0.5\textwidth]{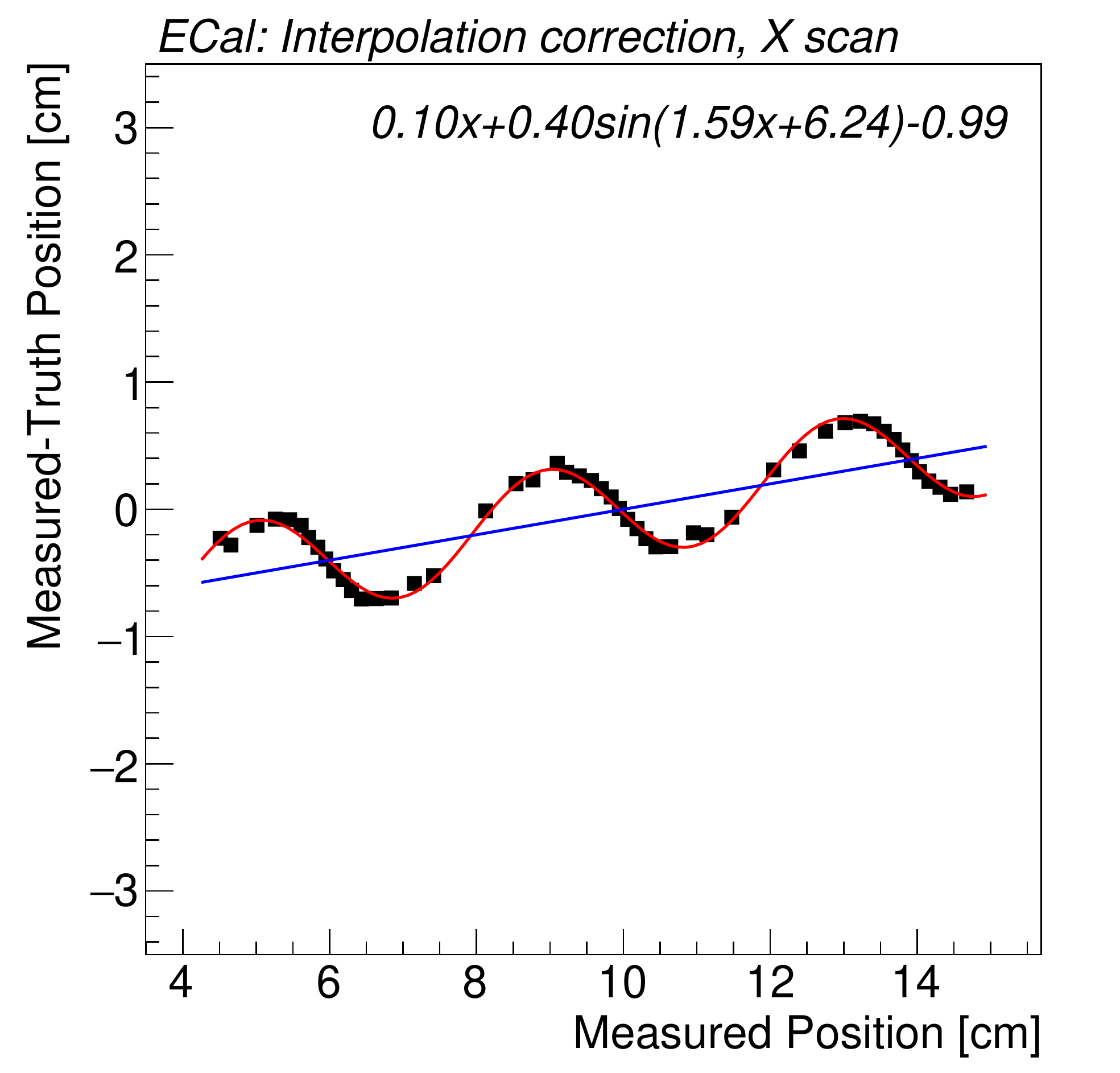}
	\caption{The systematic error of CCOG. The bias with respect to the measured position has a periodic trend, which is shown with the black squares. The fitted curve of this trend are drawn in red and its expression is presented on the top right. The blue straight line is the linear part of the function. }
	\label{fig:bias}
\end{figure*}

Charge center of gravity(CCOG) is a simple but effective method to estimate the position of the incident particles, and it is therefore extensively used in scientific position sensitive detectors\cite{de2004front,llosa2009energy,drasal2011silicon}, including the electromagnetic calorimeters\cite{meschi2001electron}. Photons or electrons coming from the interaction point will produce showers in the medium of the ECal modules, leading to signal responses in a bunch of towers, which are regarded as clusters. In the experiment, when the beam is shot in the center of one tower, the size of the cluster is typically around 3 towers, while over 70\% of the energy is collected by the central tower. With the center of gravity method, the charge weighted average of the position is defined as:
\begin{equation}
\label{eq:center}
X'=\frac{\sum_i{E_iX_i}}{\sum_i{E_i}}
\end{equation}

%its estimation is unbiased only when the energy deposited in each tower has a linear correlation with the truth position.
which shows that this method is essentially a linear operation.  However, for calorimeters, the incident particles would deposit a large amount of energy in the vicinity of the interactions, while very little in other areas. The amount of energy drops rapidly with the distance, which is apparently non-linear. Therefore the estimations given by CCOG are biased and their systematic errors can be found in Fig.\ref{fig:bias}. Every black square in the plot is the average value of the bias in every scanning point, and the "sin" shape (or "S" shape if part of points are considered) formed by these squares should be corrected. It can be described by the sum of a trigonometric and a linear function:
%Since the position scanning in this experiment is actually one dimensional at a time, so only the results in the X direction are discussed in this part. 
\begin{equation}
\label{eq:fit}
f(x)=a_0x+a_1sin(a_2x+a_3)+a_4
\end{equation}

In Fig.\ref{fig:bias}, the black squares are fitted with Eq.\ref{eq:fit}, and the fitting curve is drawn in red. The blue straight line in this plot is the linear part of the function, and those undetermined coefficients are displayed on the top right of the plot. For the trigonometric part, the peak and valley correspond to the beam shot at 1/4 and 3/4 of the tower respectively, while the center points correspond to the center or edges of a tower. This illustrates that the measured position is larger than the truth at all the 1/4 regions, and smaller at 3/4, which means the estimated value of CCOG is biased towards the center of the tower. The linear part of the fitting function is increasing with the measured position. This is caused by the geometry of the trapezoid tower shown in Fig.\ref{fig:geo}. Taking run No.3 as an example, the beam is right in the middle of the 3rd rectangular tower, but in between the 3rd and the 4th for the trapezoid geometry. The pitch of the tower used to calculate the estimation with Eq.\ref{eq:center} is the same as the rectangular geometry, resulting in an upward trend of the systematic error from run1 to run4. 

\begin{figure}
    \centering
    \begin{subfigure}[b]{0.45\textwidth}
        \includegraphics[width=\textwidth]{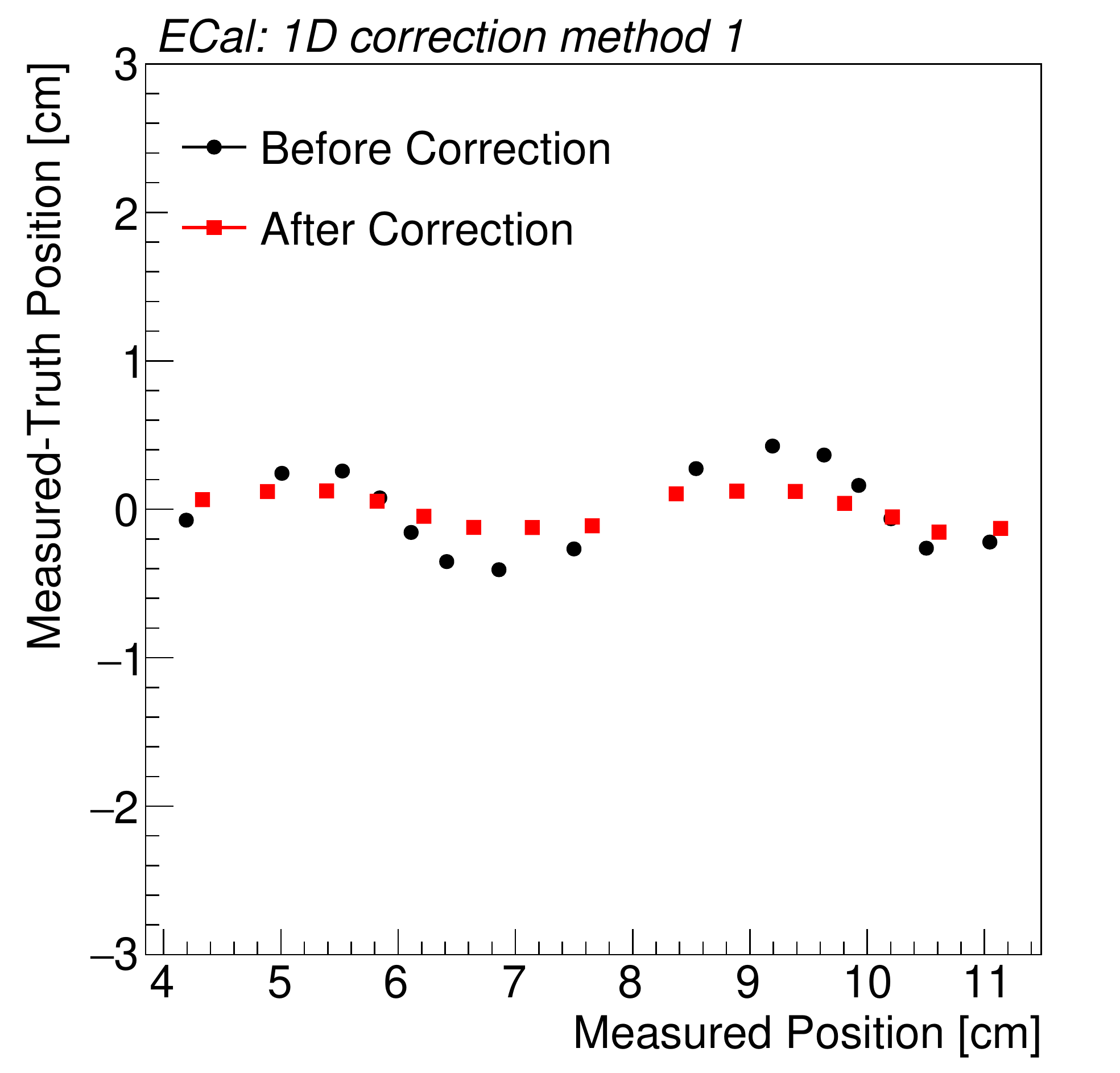}
        \caption{Corrected bias of method 1.}
        \label{fig:cor1}
    \end{subfigure}
    \begin{subfigure}[b]{0.45\textwidth}
        \includegraphics[width=\textwidth]{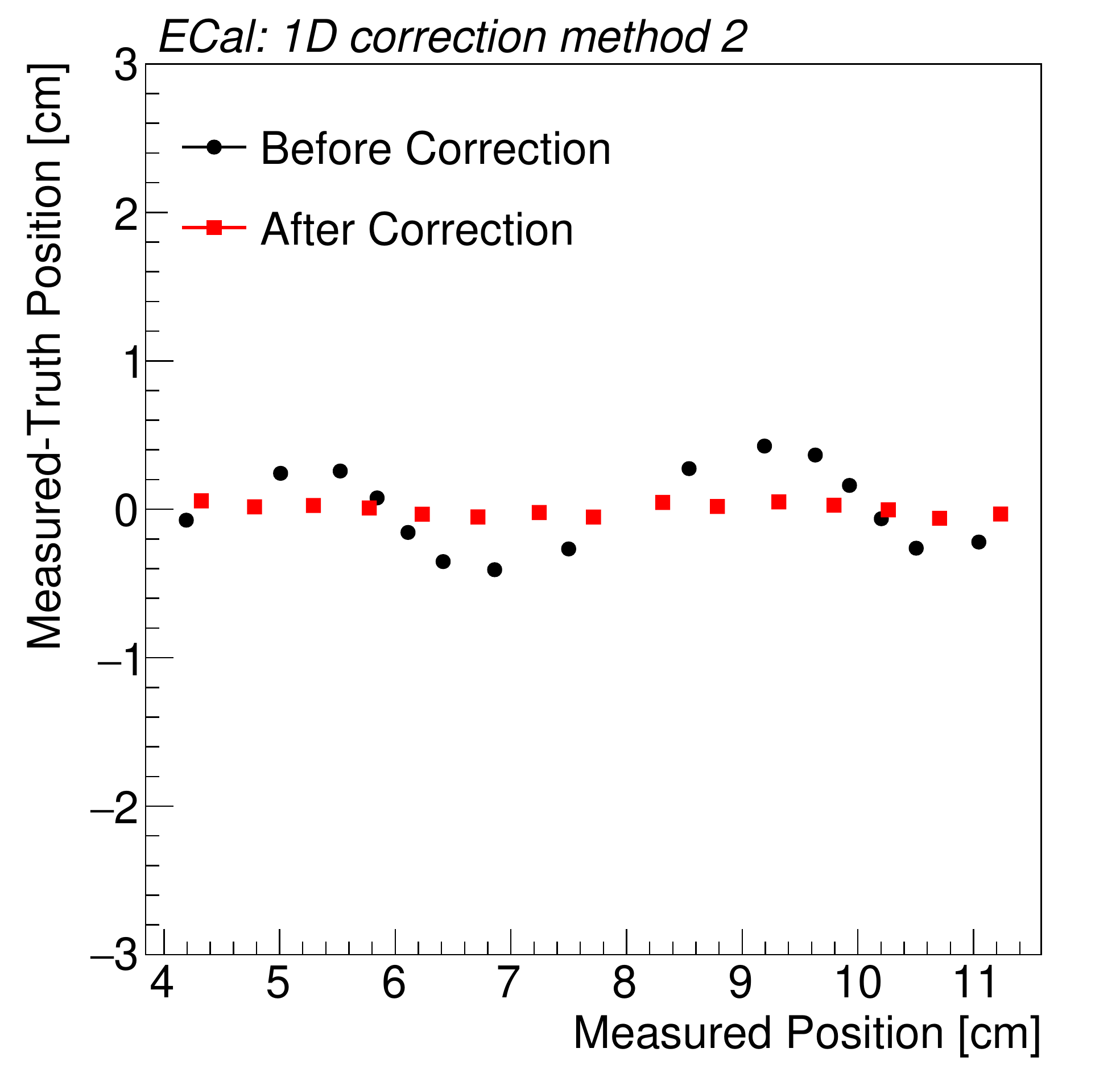}
        \caption{Corrected bias of method 2.}
        \label{fig:cor2}
    \end{subfigure}
    \caption{The comparison of systematic error before and after the correction with method 1(a) and 2(b).}
    \label{fig:1Dcor}
\end{figure}

Corrections of the position based on the fitting curve are usually applied to remove the bias and achieve good precision. Fig.\ref{fig:1Dcor} shows the bias with respect to the measured position after two 1 dimensional corrections. Correction method 1 corrects the error by fitting it with Eq.\ref{eq:fit} and subtracts the corresponding function value for every data point. After the correction, the "sin" shape is weakened but not fully eliminated, and the results are demonstrated in Fig.\ref{fig:cor1} with the red squares. So method 2 that uses a larger $a_1$ is applied to better eliminate the bias, and the results are revealed in Fig.\ref{fig:cor2}. Although method 2 has an advantage over 1 in correcting the mean of the bias, its position resolution is not as good as 1. This is presented and explained in Sec.\ref{sec:results}.

\begin{figure}
    \centering
    \begin{subfigure}[b]{0.45\textwidth}
        \includegraphics[width=\textwidth]{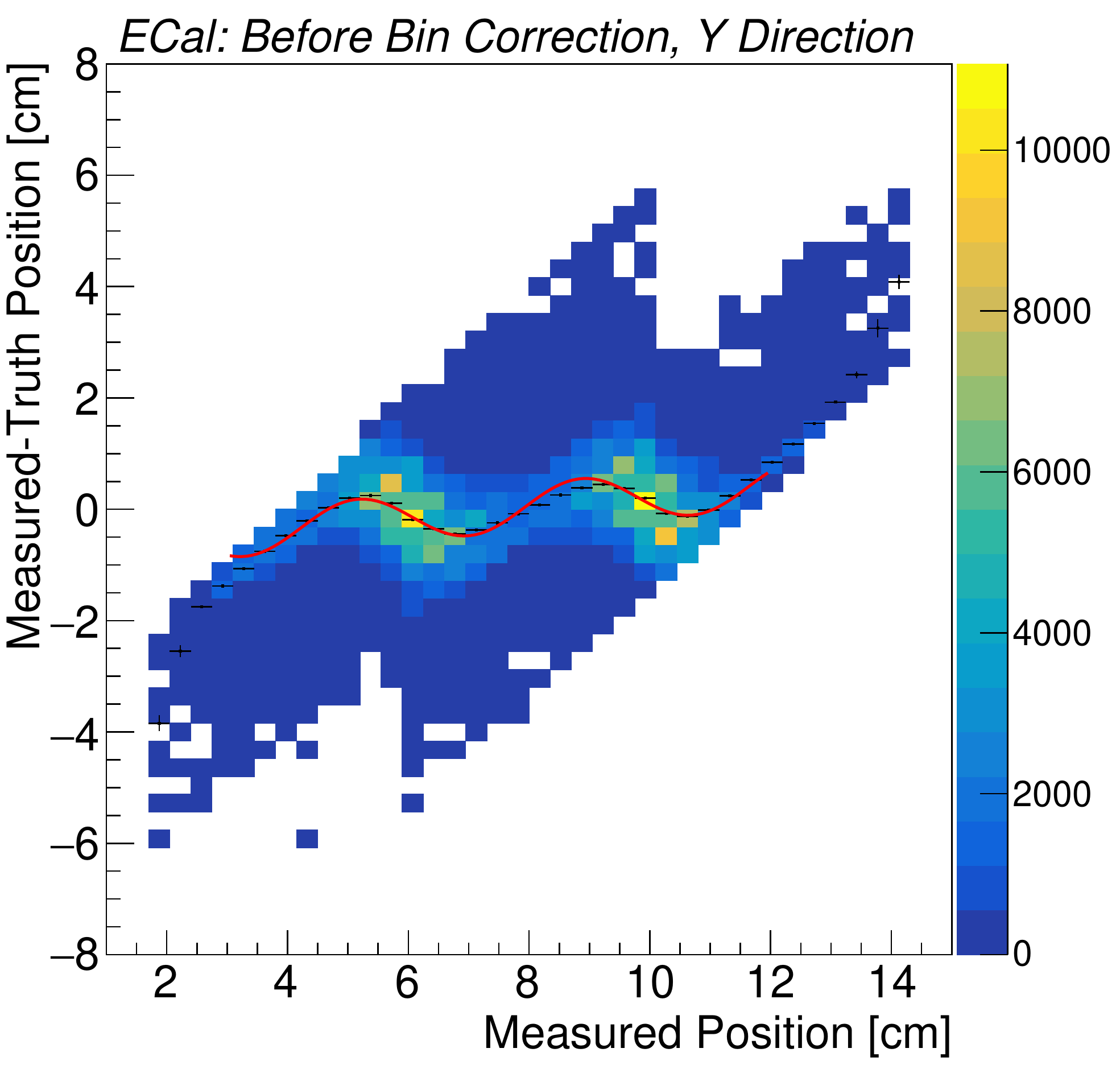}
        \caption{Before correction}
        \label{fig:be2D}
    \end{subfigure}
    \begin{subfigure}[b]{0.45\textwidth}
        \includegraphics[width=\textwidth]{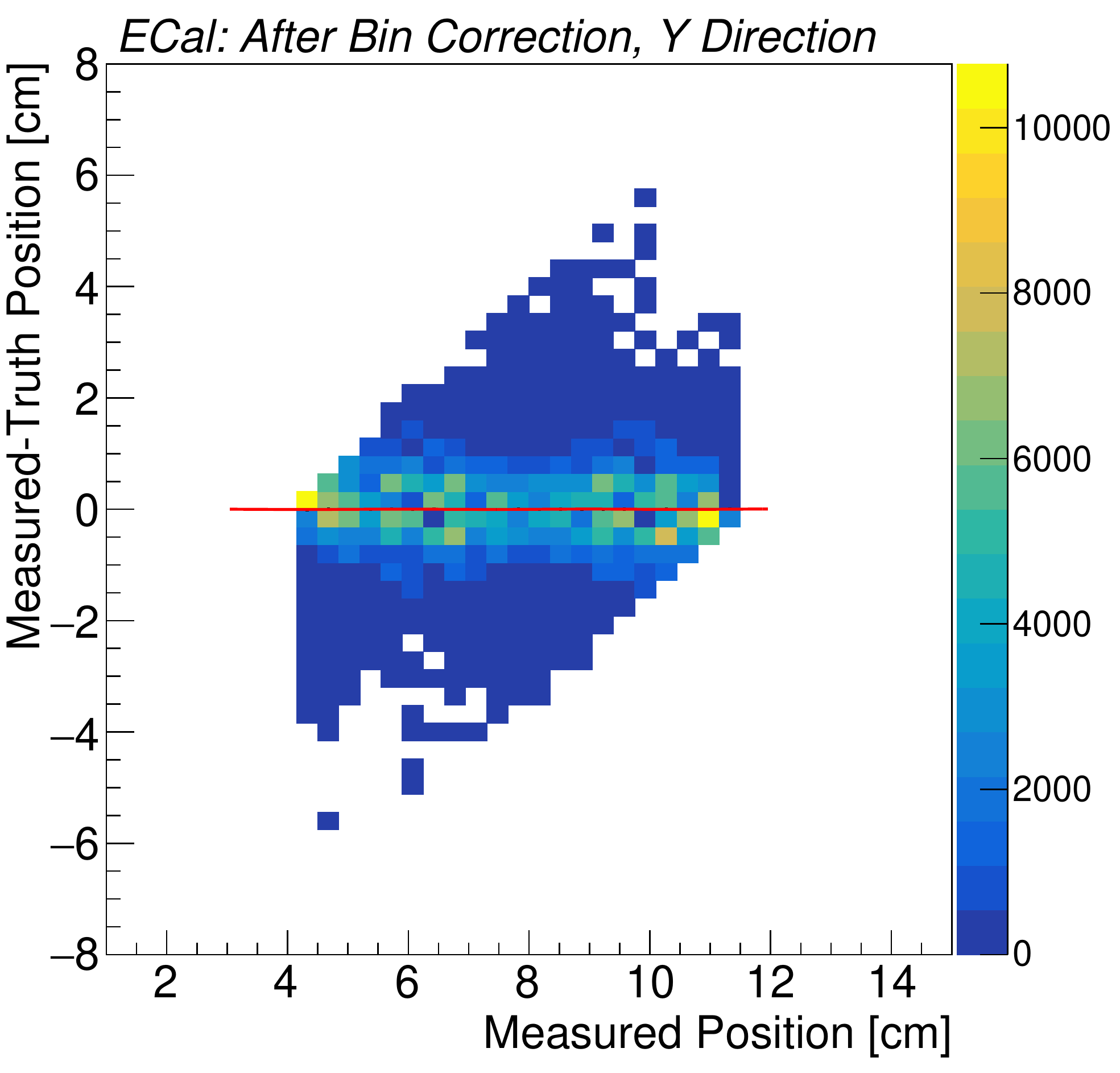}
        \caption{After correction}
        \label{fig:af2D}
    \end{subfigure}
    \caption{2D distribution of the bias and the measured position. (a) is the distribution before the bin correction, while (b) after the correction}
    \label{fig:2Dcor}
\end{figure}

In this paper, an iterative "bin correction" is also proposed. The measured positions of CCOG are filled into a two-dimensional histogram with very small bin width. The "sin" shape between the bias and the measured position is apparent in Fig.\ref{fig:be2D}. The positions in every "Measured Position" bin are corrected by subtracting the mean of the bias in this bin, and this process is iterated multiple times until the periodic trend is finally eliminated. Fig.\ref{fig:af2D} shows the histogram after all the iterations of the correction. The performance of bin correction is supposed to be better than the above two, not only because it is iterated for several times, but also because the bin width is defined by the users and it can be much smaller than the step size of the position scanning in the experiment. Both of these operations enable a more accurate characterization of the relationships in the data.

\subsection{Pulse shape analysis with neural networks}
\label{sec:NN}
Since only the total energy deposition in the related towers is used to reconstruct the position, it can be regarded as a simple solution. In order to further improve the position resolution, it is also necessary to explore some other reconstruction methods. In recent years, deep learning and neural networks have successfully solved a lot of nonlinear problems and have been widely used to reconstruct particles in various high energy physics detectors\cite{wang2018study,atlas2014neural,wang2018impact}. As early as 1998, CMS group has tried to apply fully-connected neural networks to reconstruct the position in the ECal detectors\cite{daskalakis1998monte}. However, due to the simple structure of the network and the difficulty of the training, the performance of this new method is just as good as the traditional charge center of gravity. This paper designs a more general and powerful convolutional neural network (CNN), trains and validates the network with recently developed hardware and technologies. The performance is indeed better than the center of gravity.

The original signals read out by the electronics of every event are waveforms from a collection of neighboring towers. For every event, the tower with the largest signal is considered to be the central tower. Waveforms from the central tower and its 8 nearest neighbors ($3\times3$) are selected and fed into the neural network, because they convey position distribution of the energy loss, which is the most relevant information. The truth position of every event is used as the label in the neural network and is unified in the coordinate with respect to the central tower.
\begin{figure*}[h!]
	\centering
	\includegraphics[width=0.8\textwidth]{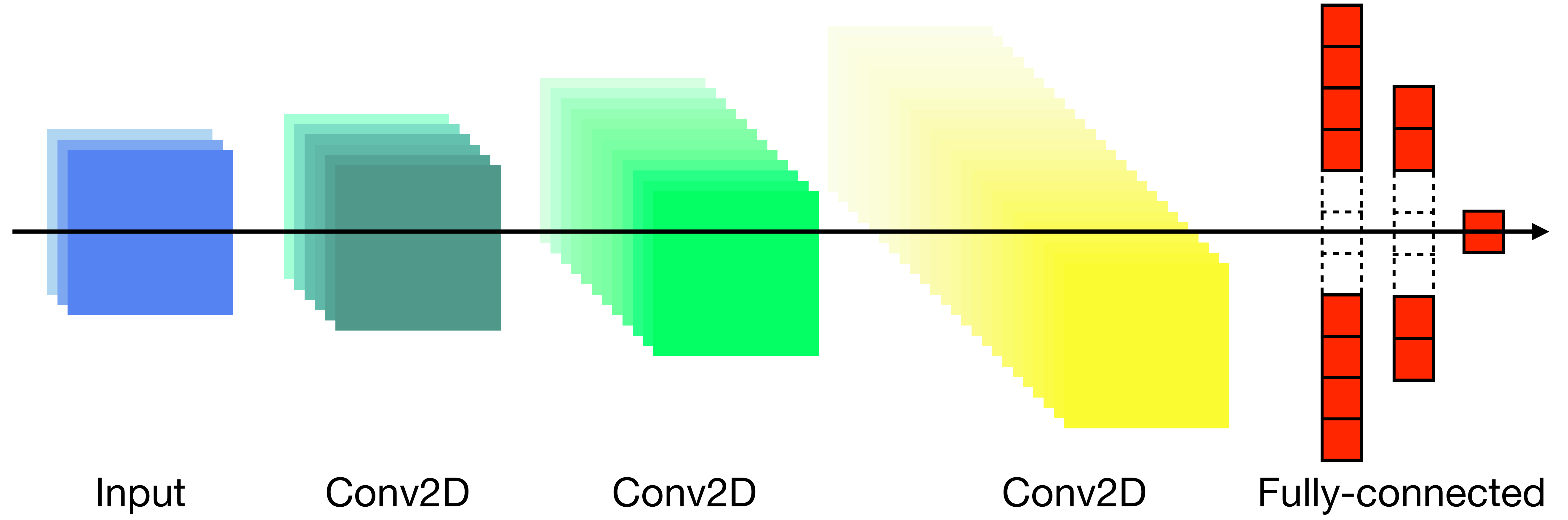}
	\caption{The structure of the convolutional neural network. The first layer is the input which is the original waveforms of the related towers. The middle 3 layers denoted as "Conv2D" are CNN layers with a same kernel size of 2. The last few layers are fully-connected layers.}
	\label{fig:CNNstructure}
\end{figure*} 

\begin{figure*}[h!]
	\centering
	\includegraphics[width=0.5\textwidth]{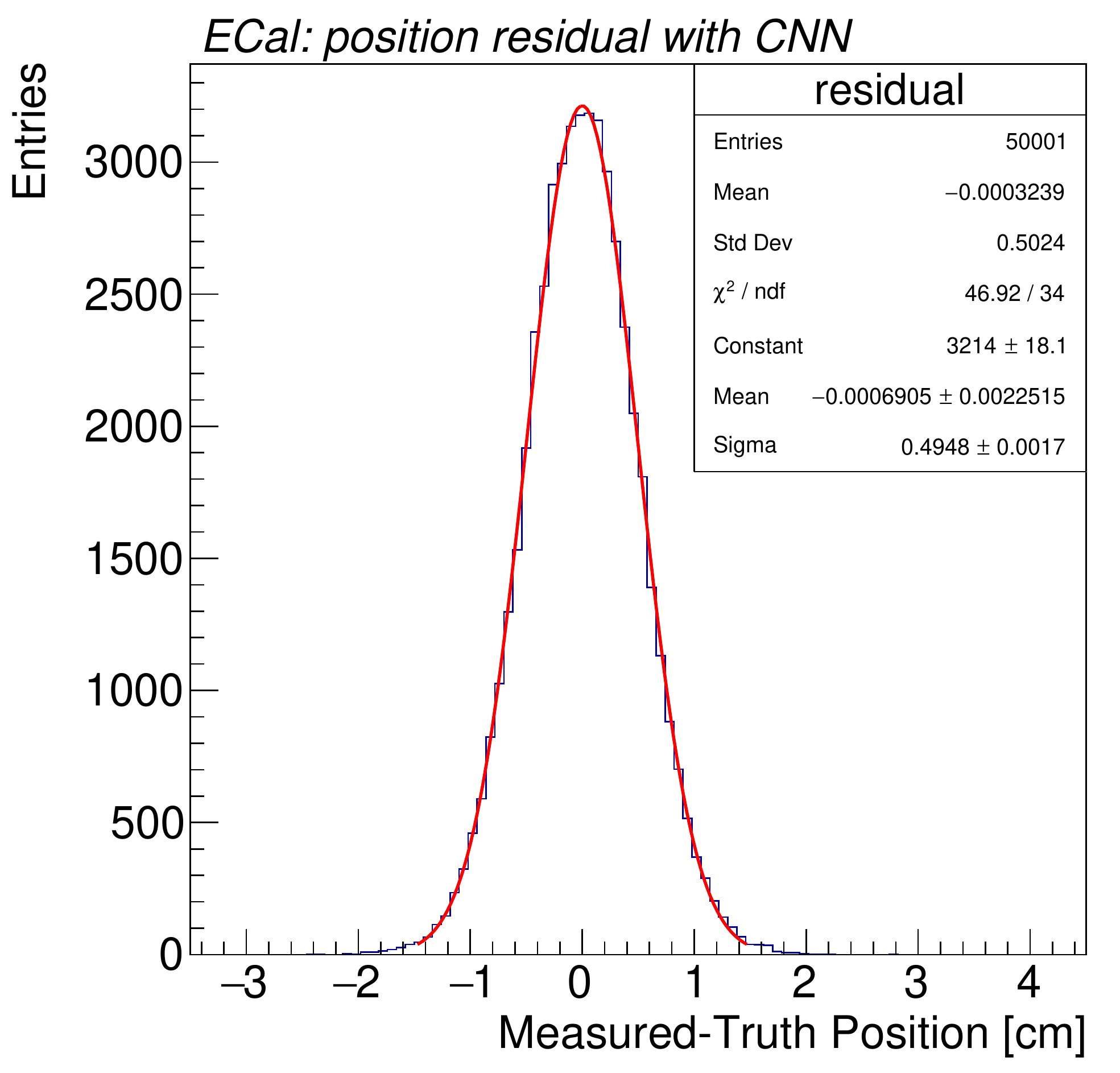}
	\caption{Distribution of the position residual. The mean value of the residual is almost 0.}
	\label{fig:CNNresi}
\end{figure*}

The structure of the network is shown in Fig.\ref{fig:CNNstructure}. It has an input layer, 3 convolutional layers, 2 fully-connected layers and an output layer. Data in the input layer has 3 dimensions, and their sizes are (9,9,3) respectively. 27 points in the middle of every waveform read out by ADC are extracted and then divided into 3 parts, each of them with 9 points. For every event, the corresponding parts in the waveforms from $3\times3=9$ towers are connected and reshaped into size $9\times9$, so that similar information of the waveforms are arranged nearby. All the 3 parts are then combined together as the network input. The middle 3 layers denoted as "Conv2D" are CNN layers with a kernel size of 2. Their dimensions are (7,7,10), (5,5,24), (3,3,64) respectively. Fully-connected layers in the end flatten the data and the final output is transformed into a single value, namely the truth position. Dropout is applied to all the middle layers to avoid overfitting. The sizes of the training and validating data are 300,000 and 100,000. The loss is the mean squared error between the estimation and the truth, and it converges very quickly in the training. Fig.\ref{fig:CNNresi} shows the distribution of the position residual estimated by the CNN network. A gaussian function is fit to the distribution within 3$\sigma$ and drawn in red. The mean value of the bias is almost 0, which illustrates that CNN is an unbiased estimation of the position.

\section{Results}
\label{sec:results}
\begin{figure}
    \centering
    \begin{subfigure}[b]{0.45\textwidth}
        \includegraphics[width=\textwidth]{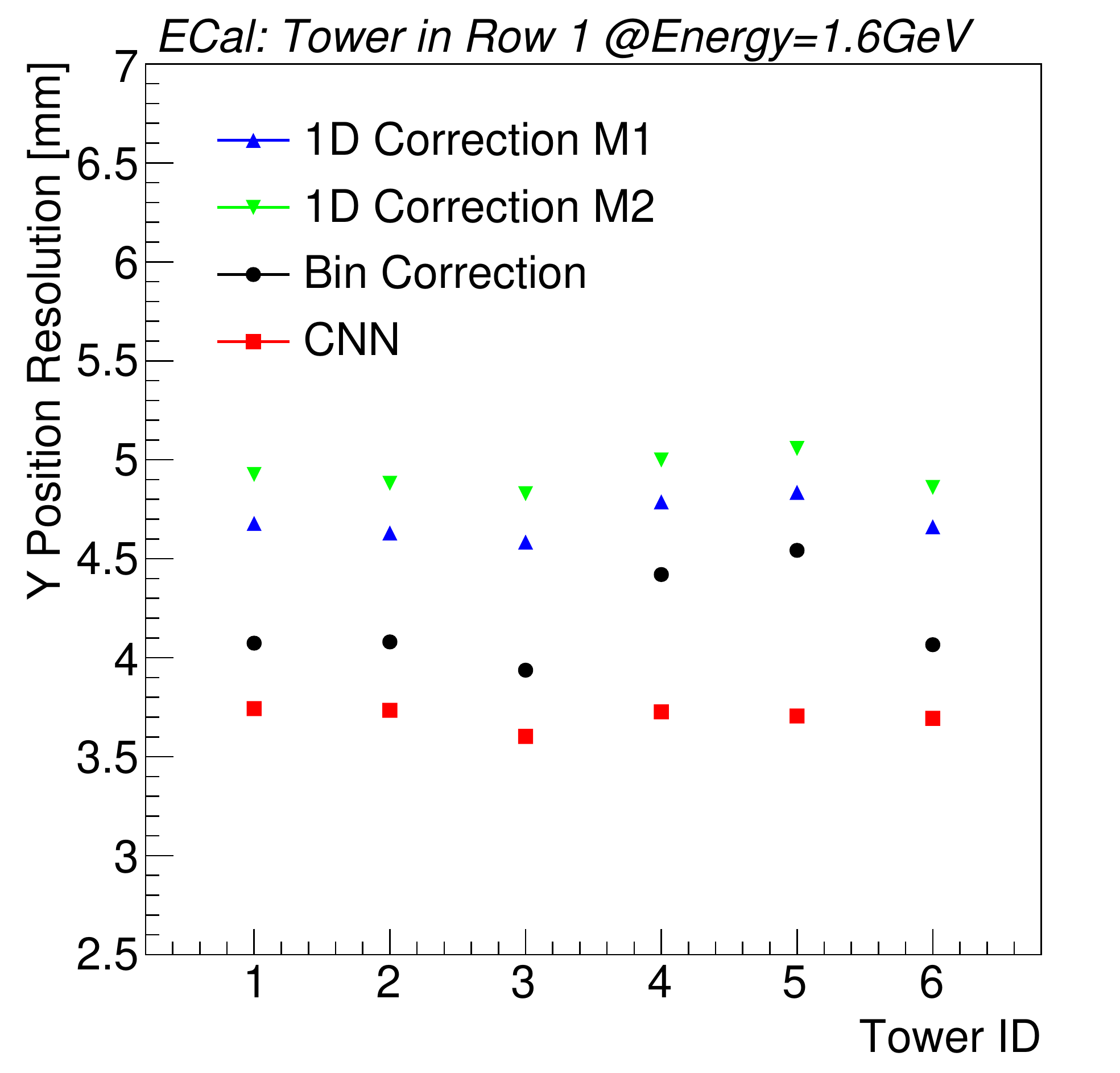}
        \caption{Row 1}
        \label{fig:yrow1}
    \end{subfigure}
    \begin{subfigure}[b]{0.45\textwidth}
        \includegraphics[width=\textwidth]{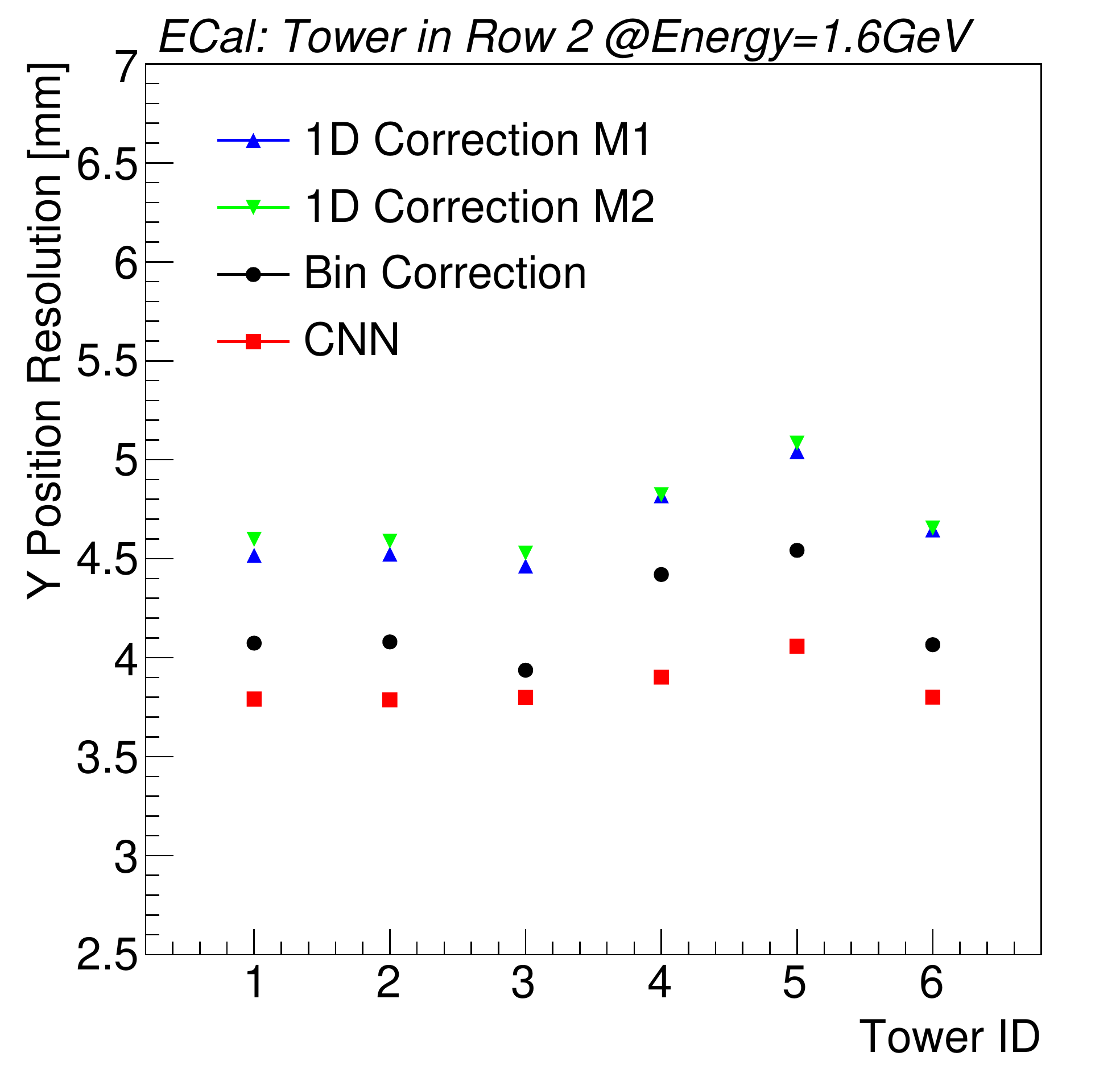}
        \caption{Row 2}
        \label{fig:yrow2}
    \end{subfigure}
    \caption{Position resolution in the Y direction. (a) shows 6 towers in row 1, while (b) in row 2. Different markers represent the results of different reconstruction methods.}
    \label{fig:yreso}
\end{figure}

The position resolution in this paper is defined to be the standard deviation of the gaussian fit to the distribution of residuals, restricted to $\pm3\sigma$. Fig.\ref{fig:yreso} shows the Y direction position resolution of the towers in 2 different rows. Different markers in the plot are the results estimated by different reconstruction methods.  The resolution is from 3.6 to 5 mm for nearly all the towers scanned in the experiment, except the 5th tower in row 2. This is because the readout SiPM of this channel was not working properly in the experiment, which also affects the nearby towers slightly. From these 2 plots, it is clear that the resolution of method 1 is smaller than 2, even if method 2 is better considering only the average of the bias in every scanning point, which is shown in Fig.\ref{fig:1Dcor}. This implies that the average of the bias should not be the only focus. It is not only because the outliers would affect the mean value, but also because the distribution of the bias itself is an important feature for judging whether the correction is effective. Method 2 which fits the data with a larger trigonometric function changes the distribution of the measured position in each scanning point, and thus leads to a worse resolution.

Compared to two 1D correction methods, bin correction improves the position resolution for all the towers shown in Fig.\ref{fig:yreso}. This demonstrates that the bin-wise correction makes the distribution of the residual more reasonable. It is also worth noting that the results obtained with CNN are even better than all the other corrections with CCOG. The position resolution is improved to only 3.8 $\rm mm$ from around 4.6 $\rm mm$, which is obtained with the widely used correction method 1. From an algorithmic point of view, charge center of gravity plus correction is just one of many effective models for the position estimation. However, the neural network is essentially a collection of models. Training the network is actually finding the most appropriate model from the collection. It is thus more generic and can propose better strategies based on the characteristics of the data for the specific problems. Besides the CNN, a fully-connected network with about 8 layers, each having around 30 nodes, was also tested. However, the resolution obtained with this network is around 4.5 $\rm mm$, so they are not shown in the paper.

\begin{figure*}[h!]
	\centering
	\includegraphics[width=0.5\textwidth]{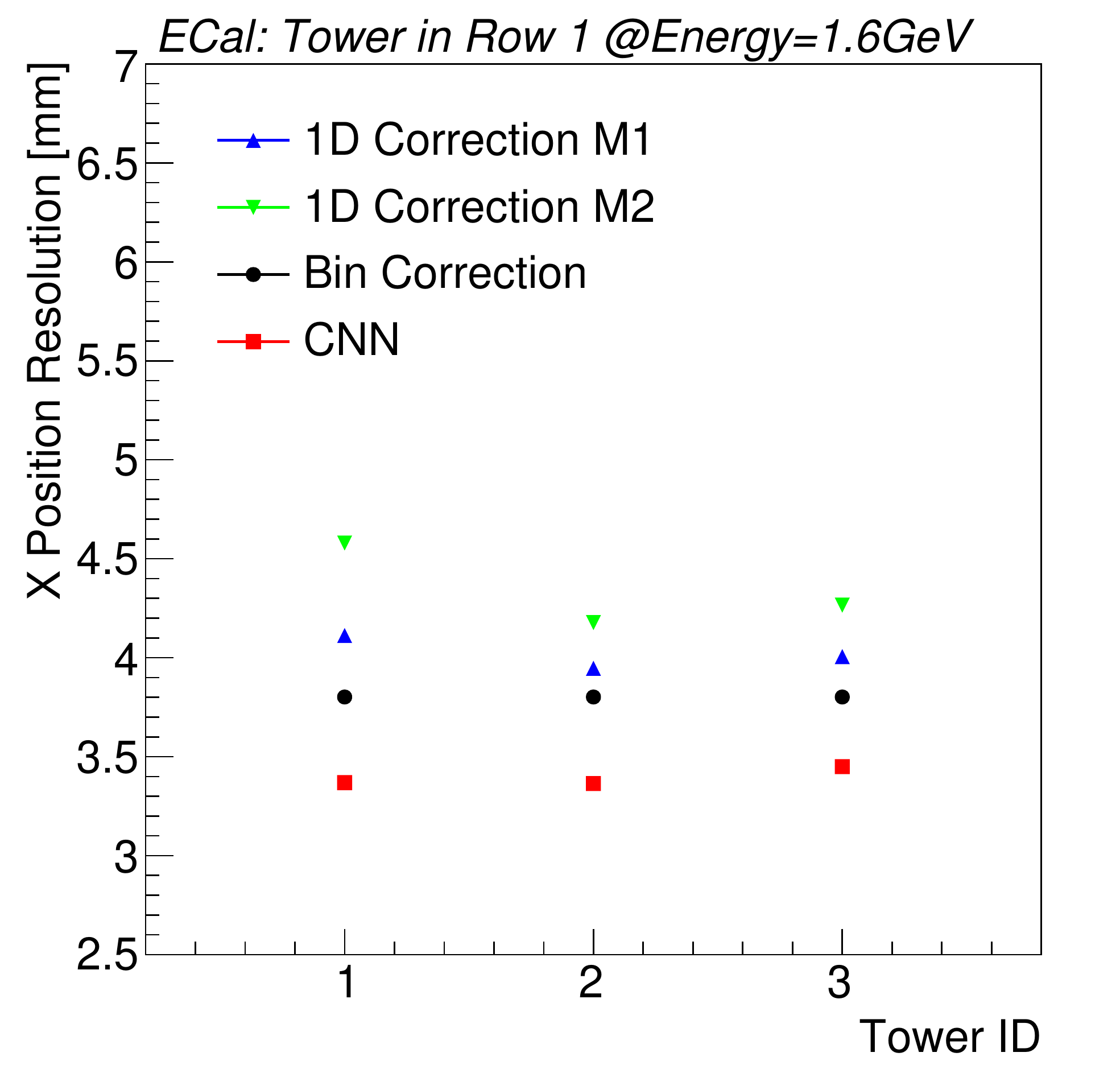}
	\caption{Position resolution in the X direction. Different markers are from different reconstruction methods.}
	\label{fig:xreso}
\end{figure*}

Fig.\ref{fig:xreso} shows the resolution in the X direction. The position scanning step along X is only 2 mm, which is less than 5 mm in the Y direction, so the scanning only spans 3 towers. It is the same that two rows of towers are scanned in the experiment, but since the resolutions are similar, only the results of the 1st row are displayed in the plot. The position resolution in the X direction is slightly better than Y, because data of a smaller scanning step would provide more detailed information on how the energy deposited in towers and its relationship with the particle position. With this smaller steps, both the methods of CCOG and CNN can extract more information, thus giving a more accurate estimation.

\begin{figure*}[h!]
	\centering
	\includegraphics[width=0.5\textwidth]{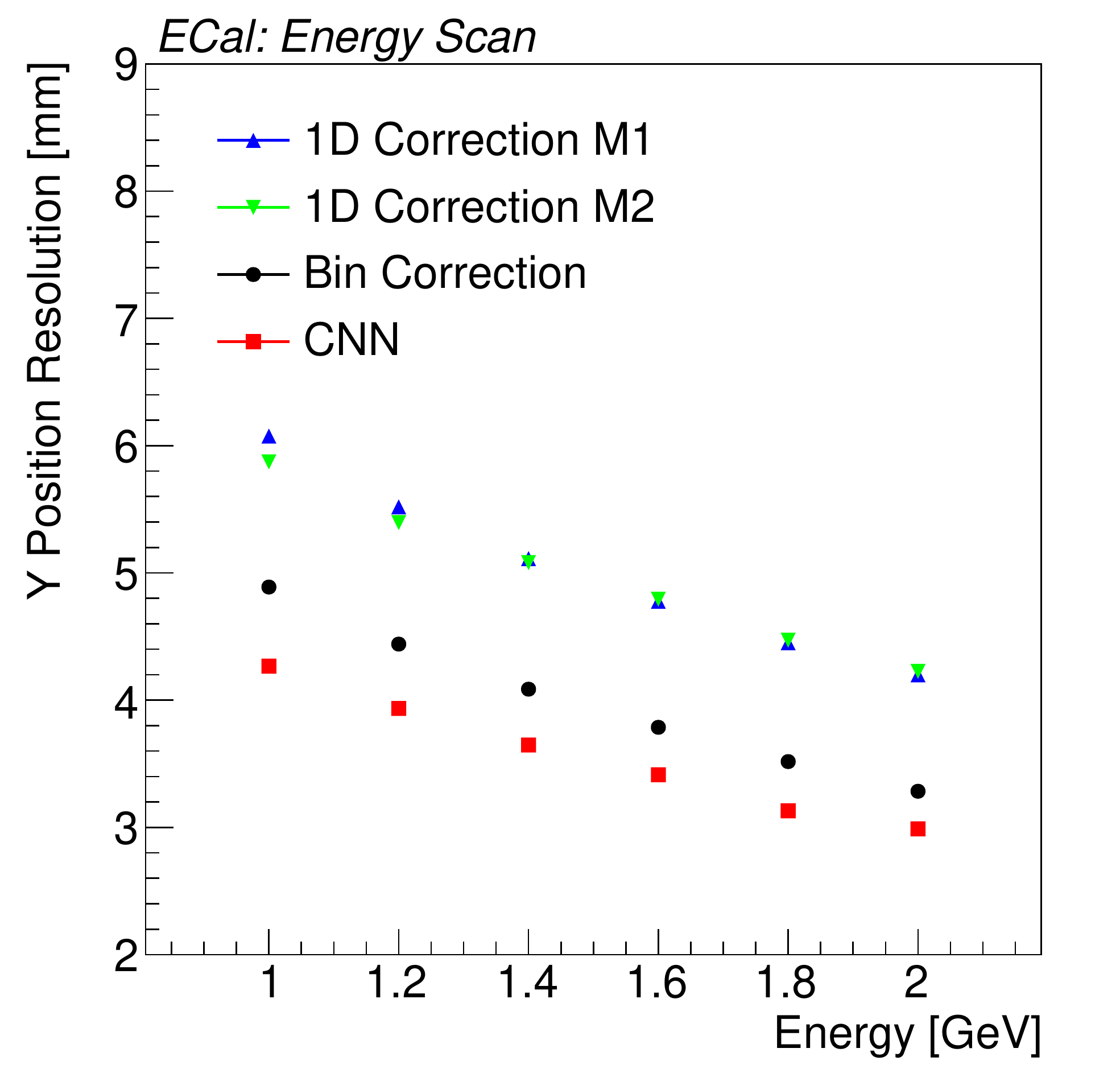}
	\caption{Position resolution with respect to the beam energy}
	\label{fig:energy}
\end{figure*}

All the position resolutions shown above are obtained with a fixed beam energy of 1.6 GeV. However, the resolution is highly related with the beam energy, because it determines the distributions and fluctuations of the electromagnetic showers. These showers would strongly affect the energy deposition in towers and thus the position accuracy. Fig.\ref{fig:energy} displays the resolution results of the ECal module made in Dubna. The towers in this module have the same geometry as the Tsinghua module but will be used in the large pseudorapidity regions. It is tested using the same experiment setup. The position resolution reduces with the beam energy. Same as before, CNN achieves the smallest resolution and then the center of gravity method with bin correction and finally, two 1D corrections. The position resolutions are improved by about 30\% with the CNN compared to the correction method 1 in all the energy scans, and it is under 3 mm when the energy reaches 2 GeV.  

\section{Conclusions}
\label{sec:conclu}
This paper carefully studies the position resolution of the ECal detector designed for the MPD experiment at NICA. The analysis is based on a beam test in DESY. The reconstruction method of charge center of gravity and some additional nonlinear corrections are described in detail. A new method based on deep learning and neural networks is also proposed, which improves the position resolution by about 30\%. It has been proved by the test that the position resolution of the Tsinghua ECal prototype is around 3.5$\sim$3.8 mm for 1.6 GeV electron beam. These results meet the requirements of the MPD experiment, and are of great significance to the further study of the detector and its application to the actual experiment. 

\section{Acknowledgments}
We sincerely thank Igor Tyapkin and Vadim Kolesnikov for their great advice and help. The work is supported by National Natural Science Foundation of China under Grant No.11420101004, 11461141011, 11275108, 11735009. This work is also supported by the Ministry of Science and Technology under Grant No. 2015CB856905, 2016 YFA0400100.

\bibliographystyle{elsarticle-num}
\bibliography{myrefs.bib}{}

\end{document}